\documentclass[superscriptaddress, twocolumn]{revtex4-1}

\usepackage{amssymb,amsmath}
\usepackage{latexsym}
\usepackage{graphicx}
\usepackage{caption}
\usepackage{subcaption}

\usepackage{color}
\usepackage{graphics}
\usepackage{epsfig}
\usepackage{bm}

\newcounter{mnotecount}[section]

\begin{document}

\newcommand{\dR}{\mathbb R}
\newcommand{\dC}{\mathbb C}
\newcommand{\dZ}{\mathbb Z}
\newcommand{\id}{\mathbb I}
\newtheorem{theorem}{Theorem}
\newcommand{\ud}{\mathrm{d}}
\newcommand{\mfn}{\mathfrak{n}}

\author{Przemys{\l}aw Ma{\l}kiewicz}
\affiliation{APC, Univ Paris Diderot, CNRS/IN2P3, CEA/Irfu, Obs de Paris, Sorbonne Paris Cit\'e, France}
\affiliation{National Centre for Nuclear Research,  00-681
Warszawa, Poland}
\email{Przemyslaw.Malkiewicz@ncbj.gov.pl}

\date{\today}

\title[]{What is Dynamics in Quantum Gravity?}

\begin{abstract}
The appearance of Hamiltonian constraint in the canonical formalism for general relativity reflects the lack of a fixed external time. The dynamics of general relativistic systems can be expressed with respect to an arbitrarily chosen internal degree of freedom, the so called internal clock. We investigate the way in which the choice of internal clock determines the quantum dynamics and how much different quantum dynamics induced by different clocks are. We develop our method of comparison by extending the Hamilton-Jacobi theory of contact transformations to include a new type of transformations which transform both the canonical variables and the internal clock. We employ our method to study the quantum dynamics of the Friedmann-Lemaitre model and obtain semiclassical corrections to the classical dynamics, which depend on the choice of internal clock. For a unique quantisation map we find the abundance of inequivalent semiclassical corrections induced by quantum dynamics taking place in different internal clocks. It follows that the concepts like minimal volume, maximal curvature and the number of quantum bounces, often used to describe quantum effects in cosmological models, depend on the choice of internal clock. 
\end{abstract}

\pacs{98.80.Qc} \maketitle

\section{Introduction}

The goal of the paper is to study the concept of quantum dynamics in the context of Hamiltonian constraint systems. The motivation for this work is the fact that the canonical formalism for general relativity, the geometrodynamics \cite{adm}, involves a Hamiltonian constraint. Hamiltonian constraints play two distinct roles in canonical formalisms: (i) they generate canonical transformations which represent dynamics and (ii) they confine physically admissible states to a submanifold in the phase space. The Hamiltonian constraint of canonical relativity follows from the diffeomorphism-invariance of that theory \cite{Ku} and any fundamental physical theory that includes gravity should involve a Hamiltonian constraint. 

There are three main approaches to quantisation of dynamics of such systems (see \cite{Ish} for a clear discussion). In this paper we report some results derived within the so called reduced phase space approach. Nevertheless, we believe that our results do not reflect the peculiarity of the specific approach that we employ but rather they follow from the properties of Hamiltonian constraint systems and cannot be avoided in any approach. We focus on a specific aspect of the quantum dynamics of gravitational systems, which is called the multiple choice problem. This problem has been discussed within the Dirac approach in \cite{Ku0} and its wider context can be found e.g. in \cite{An}. In essence, the multiple choice problem concerns the ambiguity of quantum dynamics of any Hamiltonian constraint system due to the ambiguity in the choice of the internal clock. A key tool used in our investigations is the theory of pseudocanonical transformations, which was first formulated in \cite{M} and which we develop herein. Pseudocanonical transformations treat the internal clock as a coordinate which is subject to choice as much as the canonical coordinates are subject to choice by means of canonical transformations. In order to study the ``clock effect" on quantum dynamics we propose an extended quantisation procedure which can be applied to all reduced phase spaces for all possible choices of internal clocks. The defining property of the extended quantisation procedure is the assumption of a unique quantum representation of constants of motion, the so called Dirac observables, irrespectively of the choice of internal clock. Moreover, the respective quantum theories are defined in a unique Hilbert space, where they can be compared. We use the semiclassical portrait method for describing the  semiclassical-level dissimilarities between quantum dynamics obtained for different choices of internal clocks. The goal of the present paper is not to provide a definite answer to the question of the title but rather to provide its careful and pointed formulation. We believe that our formulation should contribute to discussions of the concept of quantum dynamics for gravitational systems.

The outline of the paper is as follows. In Sec. \ref{secII} we introduce the Hamiltonian constraint formalism and discuss some of its geometrical aspects which are relevant for our purposes. In Sec. \ref{secIII} we discuss in some detail the reduced phase space approach to Hamiltonian constraint systems and develop the theory of the pseudocanonical transformations which relate different reduced phase spaces. In Sec. \ref{secIV} we apply the introduced theory in considerations of the Friedmann-Lemaitre (F-L) model of the universe. A quantisation of this model which replaces the classical singularity with a bounce is introduced in Sec. \ref{secV}. In Sec. \ref{secVI} we employ a useful method of phase space portrait to describe the quantum dynamics of the cosmological model at the semiclassical level. Sec. \ref{secVII} describes the main result of the paper. We discuss in general terms the so called extend quantisation procedure and how different quantum dynamics obtained from different reduced phase spaces can be compared. Next, we demonstrate via several examples how the redefinition of internal clock affects the quantum dynamics of the Friedmann-Lemaitre model. We conclude in Sec. \ref{secVIII}.

\section{Preliminaries}
\label{secII}

Throughout  the paper we will work with finite-dimensional phase spaces. Let us denote them by $(q_i,p_i)$. The dynamics of a Hamiltonian constraint system follows from Hamilton's equations and the constraint equation:
\begin{equation}\label{intro}
\frac{\ud ~}{\ud \tau}O(q_i,p_i)=\{O,C\},~~C=0,
\end{equation} 
where $C$ is the Hamiltonian constraint, $\tau$ is an evolution parameter and $O(q_i,p_i)$ is an observable. The motion which occurs outside the hypersurface, $C=0$, is unphysical. The multiplication of the Hamiltonian constraint $C$ by any non-vanishing function $N(q_i,p_i,\tau)$ leads to another set of equations, namely:
\begin{equation}
\frac{\ud ~}{\ud \tau'}O(q_i,p_i)=\{O,NC\},~~NC=0,
\end{equation}
which is equivalent to the former one (\ref{intro}) upon rescaling the evolution parameter:
\begin{align}\ud\tau=N\ud\tau',\end{align}
where the derivative is understood along any physical curve in $(q_i,p_i)$. This property is called the time-reparametrisation invariance and it reflects the physical assumption about the lack of a fixed external time and the auxiliary nature of the evolution parameters $\tau$ and $\tau'$.

The canonical formalism which consists of the phase space $(q_i,p_i)$ equipped with the symplectic form $\omega=\sum_i\ud q_i\ud p_i$ and a constraint function $C(q_i,p_i)$ can be reduced as follows. Let us denote the constraint surface by 
\begin{align}\mathcal{S}=\{(q_i,p_i):~C(q_i,p_i)=0\}.\end{align}  
$\mathcal{S}$ is equipped with the degenerate two-form, 
\begin{align}\omega|_{\mathcal{S}}=\sum_i\ud q_i\ud p_i|_{\mathcal{S}},
\end{align}
induced from the phase space $(q_i,p_i)$. The physical motion can be now defined in terms of integral curves generated by any vector field $v_{\mathcal{S}}$ that is null with respect to the induced form, namely
\begin{equation}\label{intcur}
\frac{\ud u^{a}}{\ud \tau}=v_{\mathcal{S}}(u),~a=1,\dots, 2n+1,~~~\omega|_{\mathcal{S}}(v_{\mathcal{S}})=0,\end{equation}
where $\{u^a\}$ is a coordinate patch in $\mathcal{S}$. The freedom in parametrising the physical motions is encoded in the ambiguity of $v_{\mathcal{S}}$ as 
\begin{align}\omega|_{\mathcal{S}}(N\cdot v_{\mathcal{S}})=N\cdot\omega|_{\mathcal{S}}(v_{\mathcal{S}})=0,
\end{align}
where $N$ is any non-vanishing function on $\mathcal{S}$. The Hamiltonian formulation of dynamics on $\mathcal{S}$ will be considered in the next section. 

\begin{figure}[t]
\centering
\includegraphics[width=0.35\textwidth]{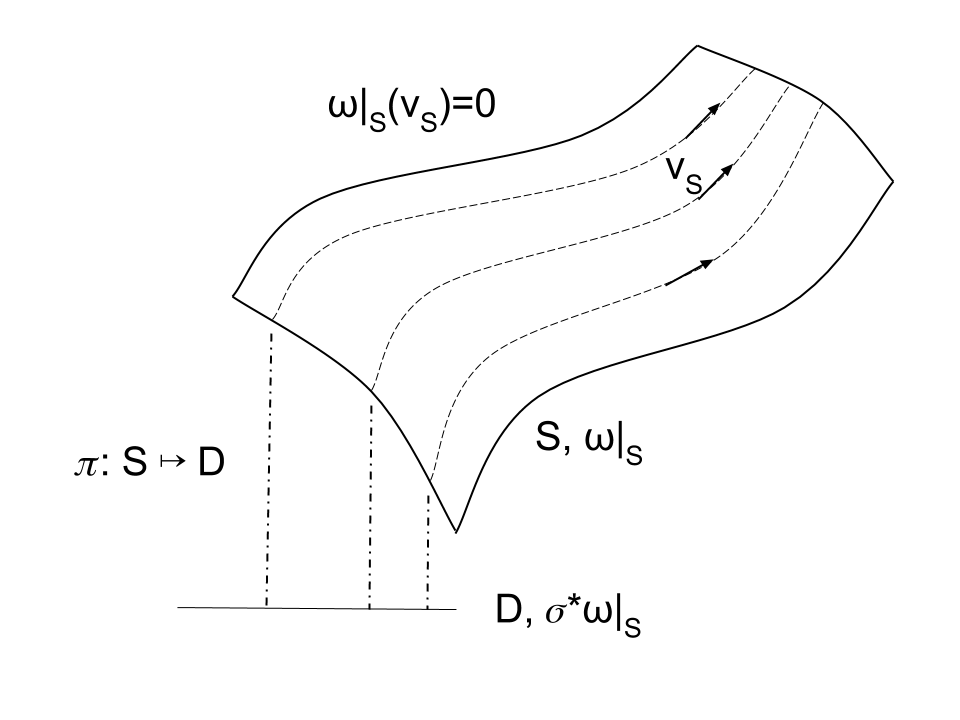}
\caption{The constraint surface $\mathcal{S}$ is given by the condition $C=0$. The Hamiltonian vector field $v_C=\{\cdot,C\}$ is tangential to that surface and generates the physical trajectories. The form $\omega|_{\mathcal{S}}$ induced from the symplectic form $\omega$ is degenerate and $v_C$ spans its null subspace.}
\label{constraint}
\end{figure}

Let us denote the space of all physical motions represented by the integral curves (\ref{intcur}) by $\mathcal{D}$ and call it the space of Dirac observables. The projection from $\mathcal{S}$ to $\mathcal{D}$ is naturally generated by the vector field $v_{\mathcal{S}}$, 
\begin{align}
\pi_{v_{\mathcal{S}}}:\mathcal{S}\mapsto \mathcal{D},
\end{align}
where any two points in $\mathcal{S}$ connected by a curve (\ref{intcur}) are mapped to a single point in $\mathcal{D}$. Consider a section,
\begin{align}
\sigma: \mathcal{D}\mapsto \mathcal{S},
\end{align} 
which is an injective map such that $\pi_{v_{\mathcal{S}}}\circ\sigma=Id$. It can be shown that any section defines a unique, non-degenerate form, $\omega|_{\mathcal{D}}$ in $\mathcal{D}$,
\begin{align}\omega|_{\mathcal{D}}:=\sigma^*(\omega|_{\mathcal{S}}),
\end{align}
which is induced from $\omega|_{\mathcal{S}}$ by $\sigma$. In other words, the space of Dirac observables is a phase space equipped with an unambiguous symplectic structure descendent from $\omega$.  Note that the constraint surface $\mathcal{S}$, where the physical motion occurs, is not equipped with any symplectic form. The structure of the constraint surface is depicted in Fig. (\ref{constraint}).

Now the problem of quantisation of dynamics of a Hamiltonian constraint system can be formulated as follows. On one hand, quantisation as commonly defined is a linear map from classical observables to quantum operators, which respects the canonical structure in the space of classical observables (see e.g. \cite{DIR}). On the other hand, the dynamics can be described only by means of dynamical observables which are functions on the constraint surface $\mathcal{S}$, because this is where the physical motion, represented by the integral curves (\ref{intcur}), happens. However, $\mathcal{S}$ is not equipped with any canonical structure. The available canonical structure is present in the space of Dirac observables $\mathcal{D}$, which do not describe the dynamics. In other words, the Poisson bracket between dynamical observables is not available.

The reader could disagree by saying that one has the symplectic form $\omega$ in the original phase space $(q_i,p_i)$, which gives the Poisson bracket $\{\cdot,\cdot\}=-\omega^{-1}$ between {\it any} observables, including those on the constraint surface $\mathcal{S}$. This, however, is false. Since only the constraint surface $\mathcal{S}$ includes the physically admissible states, any observable $O(q_i,p_i)$ defined in the original phase spaces $(q_i,p_i)$ is in fact an element of the following equivalence class:
\begin{align}
[O]=\{O'(q_i,p_i):~O'(q_i,p_i)\approx O(q_i,p_i)\},
\end{align} 
where `$\approx$' means `equal on the constraint surface $\mathcal{S}$'. These equivalence classes represent physical observables in $(q_i,p_i)$. The Poisson bracket between the equivalence classes is well-defined if and only if the evaluation of the Poisson bracket between any elements of two fixed equivalence classes gives an element of the same equivalence class. In other words, one requires that
\begin{align}\label{PBEQ}\{O_1+C,O_2\}=\{O_1,O_2\}+\{C,O_2\}&\approx \{O_1,O_2\},\end{align}
since $O_1+C\approx O_1$ and they are elements of the same equivalence class, $[O_1]$. Eq. (\ref{PBEQ}) holds if and only if 
\begin{align}\{C,O_2\}\approx 0,\end{align}
that is, $O_2$ is a (weak) constant of motion. This confirms our previous assertion that the canonical structure is unambiguously defined only for the Dirac observables, which are constants of motion. 

On one hand, quantisation of the Dirac observables seems to be (at least formally) a straightforward task. On the other hand, quantisation of dynamical observables encounters a severe obstacle in the lack of appropriate canonical structure. In the rest of the paper we wish to shed some light on this problem. In particular we will consider possible inequivalent extensions of the canonical structure on $\mathcal{D}$ to $\mathcal{S}$ and associate them with the choice of internal clock.

\section{Theory of reduced formalism}
\label{secIII}

In this section we provide a definition of the internal clock. We show how the internal clock naturally extends the canonical structure on the space of Dirac observables to the space of dynamical observables. Also, we show how the internal clock enables to reformulate the motion given by Eq. (\ref{intcur}) in terms of the Hamilton equations of a reduced canonical formalism. Then we recall the Hamilton-Jacobi theory of contact transformations. We propose an extension to this theory, which includes clock transformations (see also \cite{M}). Within this extension internal clocks can be associated with some natural canonical variables and this turns out to be a crucial tool to investigate quantisation of dynamics in  different clocks.

We assume a $2n+1$-dimensional manifold $\mathcal{S}$ equipped with a 2-form $\omega|_{\mathcal{S}}$ of rank $2n$, whose null vector fields generate a $2n$-dimensional space of curves in $\mathcal{S}$, which represent the physical motion. 

\begin{figure}[t]
\centering
\includegraphics[width=0.3\textwidth]{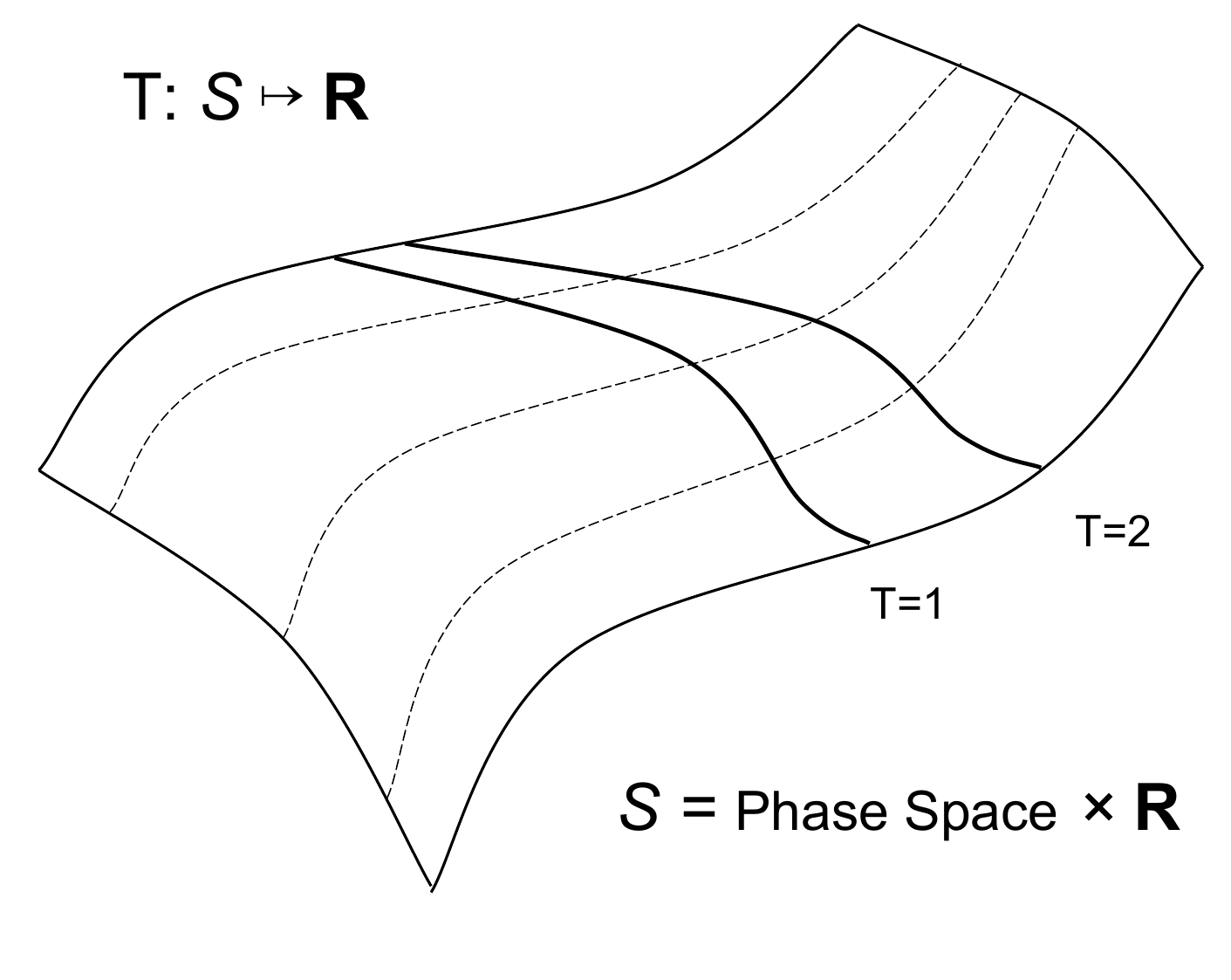}
\caption{The constraint surface $\mathcal{S}$ is sliced by level sets of the internal clock. As a result, the constraint surface is factorised into the cartesian product of the time manifold and the phase space. The trajectories are monotonic with respect to the internal clock.}
\label{clock}
\end{figure}

By internal clock, denoted by $T:\mathcal{S}\mapsto\mathbb{R}$, we mean a real function on $\mathcal{S}$ which is monotonic along each curve representing the physical motion and whose level sets cross each curve at most once, which is depicted in Fig. (\ref{clock}). The choice of the internal clock is largely arbitrary.

The restriction of the 2-form $\omega|_{\mathcal{S}}$ to a level set $T=const$ gives an invertible, and thus symplectic, 2-form $\omega|_{\mathcal{S},T=const}$. It can be inverted at each level set to obtain a Poisson bracket, 
\begin{align}\{\cdot ,\cdot\}_{T=const}=-\omega|_{\mathcal{S},T=const}^{-1}.\end{align}
Smooth repetition of this procedure for all level sets of $T$ defines a smooth Poisson bracket in the entire $\mathcal{S}$, denoted by $\{\cdot ,\cdot\}_{T}$. It can be used to compute commutation between {\it any} two observables which are functions on $\mathcal{S}$.

Let us recall the Hamilton-Jacobi theory \cite{AbMa}. The basic object in this theory is the contact manifold, $\mathcal{M}_C$. It is defined as the Cartesian product of the phase space $\mathbb{P}$ and a real time manifold, $\mathcal{M}_C=\mathbb{P}\times\mathbb{R}$. It is equipped with the contact form,
\begin{align}
\omega_{C}=\omega-\ud t\ud H,
\end{align}
where $\omega=\omega_{C}|_{t=const}=\ud q\ud p$ is a symplectic form pushed-forward from the $2n$-dimensional phase space $\mathbb{P}$ to the contact manifold $\mathcal{M}_C$, $t$ is a time parameter and $H$, called the Hamiltonian, is some function on $\mathcal{M}_C$. $\omega_{C}$ is in fact a degenerate 2-form on a $2n+1$-dimensional manifold. Its null vector fields generate curves which represent the motion in a given model. The curves can be parametrised by $t$ and obtained through the respective Hamilton equations:
\begin{align}
\frac{\ud q}{\ud t}=\{q,H\},~~\frac{\ud p}{\ud t}=\{p,H\},
\end{align}
where the Poisson bracket at any fixed time reads
\begin{align}
\{\cdot,\cdot\}:=-\omega_C|_{t=const}^{-1}=-\omega^{-1},
\end{align}
which is the (minus) inverse of the symplectic form $\omega$. This framework is useful for introduction of the notion of contact (i.e. time-dependent canonical) transformations. The contact transformations are introduced as transformations of canonical coordinates which preserve the {\it form} of the contact form, i.e.:
\begin{align}\label{contact}
(q,p,t)\mapsto (\bar{q},\bar{p},t)~~\textrm{such that}~~\omega_{C}=\ud\bar{q}\ud\bar{p}-\ud t\ud \bar{H},
\end{align}
where $\bar{H}$ is a new Hamiltonian. With it one forms new Hamilton's equations 
\begin{align}
\frac{\ud \bar{q}}{\ud t}=\{\bar{q},\bar{H}\},~~\frac{\ud \bar{p}}{\ud t}=\{\bar{p},\bar{H}\},
\end{align}
which are physically equivalent to the previous ones. Note that the time coordinate $t$ is preserved by contact transformations.

\begin{figure}[t]
\centering
\includegraphics[width=0.45\textwidth]{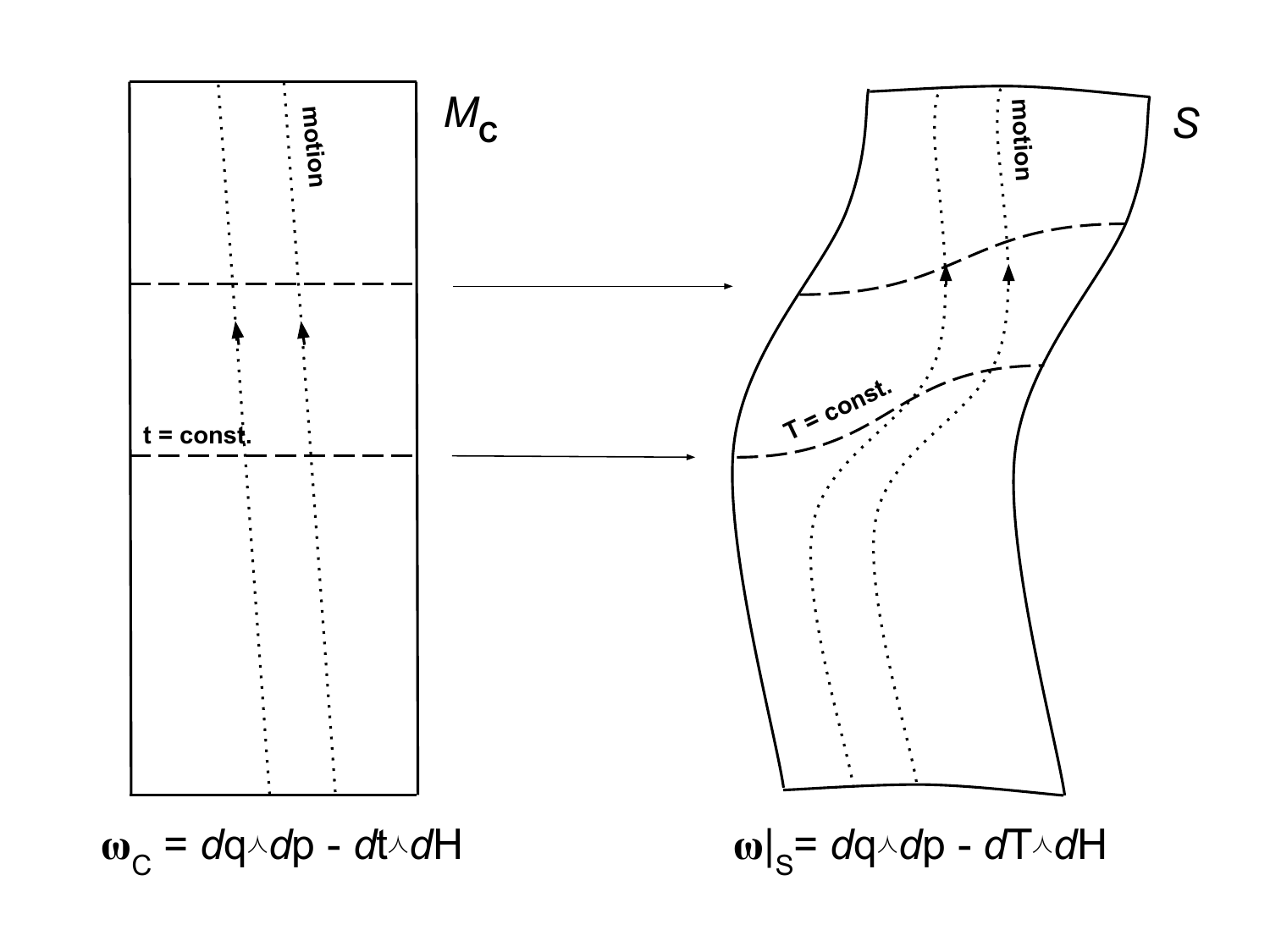}
\caption{The figure presents the identification of the constraint surface $\mathcal{S}$ (on the right) with the contact manifold (on the left) by inducing the contact coordinates $(t,q,p)$ in the constraint surface. The latter allows to rewrite the induced two-form $\omega_{\mathcal{S}}$ as the contact form with clear separation of the symplectic part, the time parameter and the Hamiltonian.}
\label{id}
\end{figure}

It is easy to see how an analogous canonical formalism can be established for the motion in the constraint surface $\mathcal{S}$ equipped with the internal clock $T$. Namely, one makes the identification of the constraint surface, the induced form and  the internal clock with the contact manifold, the contact form and  the time parameter, respectively:
\begin{align}(\mathcal{S},\omega|_{\mathcal{S}},T)\leftrightarrow(\mathcal{M}_C,\omega_C,t),\end{align}
which is depicted in Fig. (\ref{id}). We can see from the figure that the identification concerns both the curves representing the physical motion as well as time and the clock which enumerate points along each curve. Thus, it induces a one-to-one correspondence between points in $\mathcal{S}$ and points in $\mathcal{M}_C$. As a result, the canonical coordinates and the Hamiltonian on $\mathcal{M}_C$ can be mapped into the respective canonical coordinates and the respective Hamiltonian on $\mathcal{S}$. The induced form $\omega|_{\mathcal{S}}$ in the induced canonical coordinates reads:
\begin{align}
\omega|_{\mathcal{S}}=\ud q\ud p -\ud T\ud H,
\end{align}
where the internal clock $T$ plays the role of the time parameter $t$. The respective Hamilton equations,
\begin{align}\label{HamC}
\frac{\ud q}{\ud T}=\{q,H\},~~\frac{\ud p}{\ud T}=\{p,H\},
\end{align}
generate curves that are null with respect to $\omega|_{\mathcal{S}}$ and parametrised with $T$.

At this point it might look like as if we have succeeded in reducing the Hamiltonian constraint formalism to the usual unconstrained canonical formalism. This is, however, impossible. There are two important and unremovable differences:
\begin{enumerate}
  \item The internal clock is not physically the same with the time parameter. The values of the latter are irrelevant for the state of the system, which is completely characterised by the canonical coordinates in the phase space. Whereas the values of the former are indispensable for reading the state of the system. This is the reason for calling it the ``internal clock" (recall that the internal clock like all other observables is descended from the original phase space $(q_i,p_i)$).
  \item The time parameter is unique whereas there are many equally good choices for internal clock. To describe this property, the contact transformations are insufficient simply because they preserve the time coordinate as it is apparent from Eq. (\ref{contact}). Therefore, we will introduce the ``pseudocanonical transformations" to relate reduced formalisms based on different internal clocks.
\end{enumerate}

In order to include the possibility for changing the internal clock in the reduced canonical formalism we propose to extend the concept of contact transformations. We introduce the so called pseudocanonical transformations which are transformations of {\it contact} coordinates $(q,p,T)\mapsto (\bar{q},\bar{p},\bar{T})$, which include the internal clock and the pair of canonical coordinates and which preserve the {\it form} of the contact form, i.e.:
\begin{align}
(q,p,T)\mapsto (\bar{q},\bar{p},\bar{T})~~\textrm{such that}~~\omega|_{\mathcal{S}}=\ud\bar{q}\ud\bar{p}-\ud \bar{T}\ud \bar{H},
\end{align}
where $\bar{H}$ is a new Hamiltonian. The respective Hamilton equations,
\begin{align}
\frac{\ud \bar{q}}{\ud \bar{T}}=\{\bar{q},\bar{H}\},~~\frac{\ud \bar{p}}{\ud \bar{T}}=\{\bar{p},\bar{H}\},
\end{align}
are physically equivalent to Eqs (\ref{HamC}). The non-canonicity of the above transformation is the consequence of transforming the internal clock. In general, the coordinates $(\bar{q},\bar{p})$ are not canonical with respect to $T$ and $(q,p)$ are not canonical with respect to $\bar{T}$. It is easy to verify that the group of contact transformations is a normal subgroup of the group of pseudocanonical transformations, namely
\begin{align}
\forall g\in\mathcal{G}_P~~~ \mathcal{G}_Cg=g\mathcal{G}_C,
\end{align}
where $\mathcal{G}_P$ and $\mathcal{G}_C$ are the groups of pseudocanonical and contact transformations, respectively. Therefore, the group of pseudocanonical transformations is a fibre bundle $\pi:~\mathcal{G}_P\mapsto \mathcal{T}$ over the space of all internal clocks $\mathcal{T}$ with $\mathcal{G}_C$ as a fibre. The group $\mathcal{G}_P$ is regular (simple and transitive) so it can be identified with the space of all contact coordinates, $(q,p,T)$. 

Let us see how a particularly interesting class of  pseudocanonical transformations can be constructed. Suppose there are some contact coordinates $(q,p,T)$ on $\mathcal{S}$ such that 
\begin{align}
\omega|_{\mathcal{S}}=\ud q\ud p-\ud T\ud H(q,p)
\end{align}
We make a clock transformation $T\rightarrow \bar{T}=\bar{T}(q,p,T)$ and ensure that the new and the old clock are monotonic with respect to each other along the physical motion. We demand that the new canonical coordinates $(\bar{q},\bar{p})$ are such that 
\begin{align}
\omega|_{\mathcal{S}}=\ud \bar{q}\ud \bar{p}-\ud \bar{T}\ud H(\bar{q},\bar{p}),
\end{align}
that is, the Hamiltonian $H(\bar{q},\bar{p})$ is {\it formally} the same function of canonical coordinates as $H(q,p)$. One notices that in this case the Hamilton equations of motion are {\it formally} the same both in $(q,p,T)$ and in $(\bar{q},\bar{p},\bar{T})$. Thus, the constants of motion, denoted by $I_j$, can be demanded to have the same {\it form} in both contact coordinate systems, namely
\begin{align}\label{switch}
I_j(q,p,T)=I_j(\bar{q},\bar{p},\bar{T}),~~j=1,\dots,2n
\end{align}
The above set of equations supplemented with 
\begin{align}\bar{T}=\bar{T}(q,p,T)\end{align}
gives $2n+1$ algebraic relations between two coordinate systems on $\mathcal{S}$, $(q,p,T)$ and $(\bar{q},\bar{p},\bar{T})$. Given initial contact coordinates $(q,p,T)$ and a new internal clock $\bar{T}$, they can be solved for $(\bar{q},\bar{p})$ unambiguously. In geometrical terms, the above relations define a section $\sigma$ in the space of all contact coordinates:
\begin{align}
\sigma:~\mathcal{T}\ni\bar{T}\mapsto (\bar{q},\bar{p},\bar{T})\in\mathcal{G}_P.
\end{align}
This particular section will soon turn out very useful for quantising a given system in many different internal clocks.

\section{Friedmann-Lemaitre model}
\label{secIV}

Let us consider a simple gravitational model, namely the flat Friedmann-Lemaitre model of the universe $\mathbb{R}\times\Sigma=\mathbb{R}\times\mathbb{T}^3$, filled with radiation and equipped with the line element, 
\begin{align}\ud s^2=-N^2\ud t^2+a^2\delta_{ab}\ud x^a\ud x^b.\end{align}
{ Let $v_0=\int_{\Sigma}\ud^3 x$ be the coordinate volume}. It can be shown \cite{BDGM} that the Hamiltonian constraint for this model reads:
\begin{align}
C=p_T+p^2,~~(q,p)\in\mathbb{R}_+^*\times\mathbb{R},~~(T,p_T)\in\mathbb{R}^2,
\end{align}
{where $q=av_0^{\frac{1}{3}}$ and $p=\frac{3(av_0^{\frac{1}{3}})^2}{8\pi G}\frac{1}{N}\frac{\dot{a}}{a}$ are respectively the length and the expansion of the universe. Their physical dimensions are $[q]=length$ and $[p]=mass$, where we set $c=1$.} $T$ is a variable associated with the state of the fluid and $p_T$ is the respective momentum. { For fluids other than radiation, the respective Hamiltonian constraint can be formulated analogously, see \cite{BDGM}.}

\begin{figure}[t]
\centering
\includegraphics[width=0.3\textwidth]{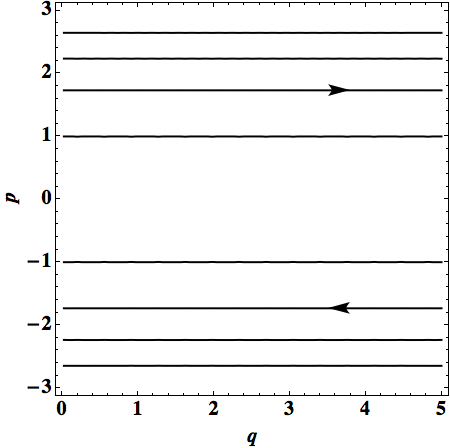}
\caption{The Hamiltonian dynamics of the flat Friedmann-Lemaitre model resembles the dynamics of a free particle on a half-line. The phase space trajectories are shown in the figure. The trajectories arriving at/starting from $q=0$ represent the big-crunch/big-bang singularity of the universe.}
\label{free}
\end{figure}

Let us first compute the induced 2-form on the surface $\mathcal{S}$, where the constraint $C=0$ vanishes:
\begin{align}\label{indFRW}
\omega|_{\mathcal{S}}=\ud q\ud p +\ud T\ud p_T\big|_{C=0}=\ud q\ud p -\ud T\ud p^2,
\end{align}
where we have chosen to parametrise the surface $\mathcal{S}$ with three independent coordinates: $q$, $p$ and $T$. Naming $H=p^2$ the reduced Hamiltonian, we deduce from the Hamilton-Jacobi theory that the motion of the system with respect to $T$ can be formulated by means of the Hamilton equations:
\begin{align}\label{hamFRW}
\frac{\ud q}{\ud T}=\frac{\partial H}{\partial p},~~\frac{\ud p}{\ud T}=-\frac{\partial H}{\partial q},~~H=p^2,
\end{align}
where the fluid variable plays the role of the internal clock and $(q,p)$ play the role of canonical coordinates in the reduced phase space. { Note that $p_T$ is removed from the formalism that now takes the form of the Hamiltonian formulation of a free particle on a half-line.} Several phase space trajectories are plotted in Fig. (\ref{free}). The phase space boundary $q=0$ represents the big-bang/big-crunch singularity.

It is straightforward to find two independent constants of motion, 
\begin{align}\label{comFRW}I_1=q-2pT,~~~I_2=p.\end{align}
Now we construct pseudocanonical transformations by following the prescription proposed in the previous section, principally in Eq. (\ref{switch}). For simplicity, we restrict our attention to the following clock transformations:
\begin{align}\label{clockFRW}
T\rightarrow \bar{T}=T+D(q,p)
\end{align}
where $D(q,p)$ is the so-called delay function and where $T$ and $\bar{T}$ are monotonic with respect to each other along the motion. Demanding 
\begin{align}q-2pT=\bar{q}-2\bar{p}\bar{T},~~p=\bar{p}\end{align}
we obtain:
\begin{align}\label{particleswitch}
\bar{T}=T+D(q,p),~~\bar{p}=p,~~\bar{q}=q+2pD(q,p).
\end{align}
The induced form (\ref{indFRW}) reads in the new variables:
\begin{align}
\omega|_{\mathcal{S}}=\ud \bar{q}\ud \bar{p} -\ud \bar{T}\ud \bar{p}^2,
\end{align}
and its form is preserved as expected. It can be easily verified that the new Hamilton equations,
\begin{align}\label{hamFRWn}
\frac{\ud \bar{q}}{\ud \bar{T}}=\frac{\partial \bar{H}}{\partial \bar{p}},~~\frac{\ud \bar{p}}{\ud \bar{T}}=-\frac{\partial \bar{H}}{\partial \bar{q}},
\end{align}
where $\bar{H}=\bar{p}^2$, define the same motion as Eq. (\ref{hamFRW}). Thus, we have succeeded in establishing the Hamiltonian formalisms of the Friedmann-Lemaitre model for {\it all} internal clocks defined by Eq. (\ref{clockFRW}). We emphasise that the canonical structures induced by different clocks are different. For instance, the computation of the Poisson bracket between $\bar{q}$ and $\bar{p}$ given in Eq. (\ref{particleswitch}) for the clock $T$,
\begin{align}\{\bar{q},\bar{p}\}|_{T=const}=\{q+2pD(q,p),p\}|_{T=const}=1+2p\partial_q D,\end{align}
gives in general a value different from 1. 

\section{Quantisation of dynamics}
\label{secV}

In what follows we quantise the dynamics of the Friedmann-Lemaitre model expressed by means of the canonical formalism,
\begin{align}
H=p^2,~~~~(q,p)\in\mathbb{R}_+^*\times\mathbb{R}.
\end{align}
Notice that $q>0$. Quantisation of the above Hamiltonian based on the representation of the affine group which is the symmetry of the half-plane was presented in \cite{BDGM}. Nevertheless, for the sake of simplicity, let us employ a simplified, though not consistent with the symmetry of the half-plane, procedure:
\begin{align}\label{qhamFRW}
H=p^2~\rightarrow~ \hat{H}\Psi(x):=-\hbar^2\triangle~\Psi(x),
\end{align}
where
\begin{align}
\Psi(x)\in C^{\infty}_c(\mathbb{R}_+)\subset L^2(\mathbb{R}_+,\ud x).
\end{align}
($C^{\infty}_c(\mathbb{R}_+)$ is the space of smooth functions with compact support on $\mathbb{R}_+$). It is known that there are infinitely many self-adjoint extensions of the above operator (see e.g. \cite{reedsimon}). We extend the domain to smooth wave-functions $\Psi\in C^{\infty}(\mathbb{R}_+)$ such that $\Psi_{,x} \in L^2(\mathbb{R}_+,\ud x)$ and satisfying the Dirichlet boundary condition, 
\begin{align}\Psi(0)=0.\end{align}
The operator is essentially self-adjoint on this domain and its spectrum is non-negative, $sp(-\triangle)=\mathbb{R}_+\cap\{0\}$.

The classical Hamiltonian $H$ that generates an incomplete motion in the phase space is promoted to the self-adjoint operator $\hat{H}$ that generates a complete unitary evolution in the Hilbert space, because the quantum dynamics as measured by the internal clock $T$ can be extended indefinitely by the virtue of the Stone-von Neumann theorem. This is how the quantisation resolves the classical singularity. We emphasise that the choice of the Dirichlet boundary condition is merely an example and there are infinitely many other conditions which ensure the self-adjointness of $\hat{H}$ (see \cite{reedsimon} for details) and lead to the singularity resolution. Instead of solving the respective Schr\"odinger equation, we will employ a semiclassical description in terms of phase space portraits.

\section{Phase space portrait}
\label{secVI}

In what follows we introduce the phase space portrait method to provide a simple representation of the unitary dynamics generated by the self-adjoint Hamiltonian $\hat{H}=-\hbar^2\triangle$ described in the previous section. We make use of coherent states based on a unitary representation of the affine group. We recall the principal elements of that framework in order to keep the presentation self-contained. { More details can be found in Appendix A and in \cite{BDGM}.}

The affine group, 
\begin{align}
(q,p)\circ (q',p')=(qq',q^{-1}p'+p),
\end{align} 
is a minimal canonical group acting in the phase space $\dR_+\times\dR$. It has a unique, up to irrelevant sign, non-trivial unitary irreducible and integrable representation, $U(q,p)$, which is defined in $L^2(\dR_+,\ud x)$ as
\begin{align}
U(q,p)[\psi(x)]=\frac{e^{-ipx/\hbar}}{\sqrt{q}}\psi\left(\frac{x}{q}\right)
\end{align}
Given a normalised vector $|\psi_0\rangle$ in the Hilbert space, the above representation can be used to define a family of the affine coherent states, $\dR_+\times\dR\ni (q,p)\mapsto |q,p\rangle\in L^2(\dR_+,\ud x)$:
\begin{align}
|q,p\rangle=U(q,p)|\psi_0\rangle,~~\langle x|q,p\rangle=\frac{e^{-ipx/\hbar}}{\sqrt{q}}\psi_0\left(\frac{x}{q}\right).
\end{align}
The fiducial state $\psi_0(x)=\langle x|\psi_0\rangle$ is admissible if $\int_{\mathbb{R}_{+}}|\psi_0(x)|\frac{\ud x}{x}<\infty$. We set 
$$\psi_0(x)=\frac{a^{a/2}}{\sqrt{\Gamma(a)}}x^{\frac{a-1}{2}}e^{-\frac{a}{2}x},$$ 
where $a>2$. { One can verify that $\langle q,p|\hat{Q}|q,p\rangle=q$ and $\langle q,p|\hat{P}|q,p\rangle=p$, where $\hat{Q}$ and $\hat{P}$ are respectively the position and momentum operators. The affine coherent state $|q,p\rangle$ is said now to represent the classical state $(q,p)$. This identification of classical and quantum states allows to make a comparison of classical and quantum dynamics as explained below.}

\begin{figure}[t]
\centering
\includegraphics[width=0.3\textwidth]{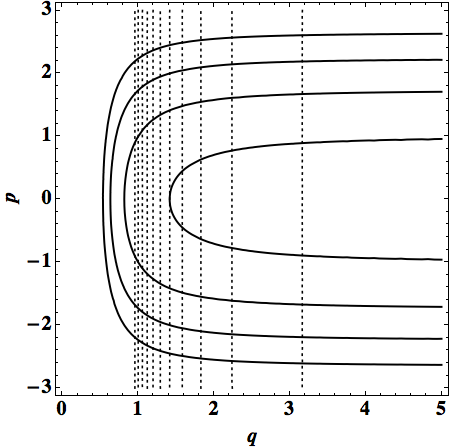}
\caption{The mostly horizontal solid lines represent the semiclassical dynamics of homogenous three-geometries of the flat Friedmann-Lemaitre universe. The vertical dotted lines depict equally-spaced values of the repulsive potential. As the universe approaches the singular state $q=0$, it is repelled by the potential, which results in a bounce. { We set $K=2$, $q$ and $p$ are given in Planck lengths and Planck masses, respectively.}}
\label{1}
\end{figure}

Let us formulate the quantum dynamics by means of the least action principle. The quantum action is defined as
\begin{align}
A_Q(\psi)=\int\langle\psi|i\hbar\frac{\partial}{\partial T}-\hat{H}|\psi\rangle~\ud t,
\end{align}
where $\psi(x)=\langle x|\psi\rangle$ is a regular function of `$x$'. It is minimised by solutions to the Schr\"odinger equation, $i\hbar\frac{\partial\psi(x)}{\partial T}=\hat{H}\psi(x)$. The quantum action $A_Q$ can be restricted to the family of coherent states \cite{klauder}: 
\begin{align}\label{semiA}
A_Q(q,p)=\int \langle q,p|i\hbar\frac{\partial}{\partial T}-\hat{H}|q,p\rangle\ \ud t
\end{align}
The phase space portrait is obtained via minimisation of the action (\ref{semiA}) with respect to $q$ and $p$. We obtain the semiclassical Hamilton equations:
\begin{align}\label{semiHE}
\frac{\ud q}{\ud T}=\frac{~\partial H^{sem}}{\partial p},~~\frac{\ud p}{\ud T}=-\frac{~\partial H^{sem}}{\partial q},
\end{align}
where $H^{sem}(q,p)=\langle q,p|\hat{H}|q,p\rangle$. In \cite{BDGM} it was found that
\begin{align}
H^{sem}(q,p)=p^2+\frac{\hbar^2K}{q^2},
\end{align}
{ where $\hbar^2K=\langle \psi_0|\hat{H}|\psi_0\rangle=\frac{\hbar^2a^2}{4(a-2)}>0$ is the expectation value of energy in the fiducial state $|\psi_0\rangle$. Eqs (\ref{semiHE}) generate an approximate motion that has an important property, namely it occurs both in the Hilbert space, $T\mapsto |q(T),p(T)\rangle$, and in the phase space, $T\mapsto (q(T),p(T))$. The motion in the phase space describes quantitatively, in terms of the classical observables $q$ and $p$, basic features of the quantum dynamics such as the fact that it is nonsingular. Notice that away from $q=0$, the term $\frac{\hbar^2K}{q^2}$ vanishes and the semiclassical dynamics becomes classical. On the other hand, close to $q=0$, the term $\frac{\hbar^2K}{q^2}$ is a steep, repulsive potential that eventually stops and reverses any particle approaching the boundary $q=0$. Hence, the singularity of the classical Friedmann-Lemaitre universe is avoided. 

In what follows we use Planck units because of the value of $\hbar^2K=\hbar^2\frac{a^2}{4(a-2)}=m_P^2l_P^2\frac{a^2}{4(a-2)}$, where $m_P$ and $l_P$ are respectively Planck mass and Planck length. The values of $q$ and $p$ are given in $l_P$ and $m_P$, respectively. Furthermore, we assume $\hbar^2K=2m_P^2l_P^2$. Several semiclassical trajectories are plotted in Fig. (\ref{1}).}

\section{Switching between clocks}
\label{secVII}

We already know from the section {\it Theory of reduced formalism} and in particular from Eq. (\ref{switch}) how to switch between reduced canonical formalisms based on different internal clocks. For the case of a free particle on the half-line we showed that the pseudocanonical transformations take the form (\ref{particleswitch}) and, indeed, preserve the formal dependence of constants of motion and of the induced two-form on the basic variables and the internal clock. In the section {\it Quantisation of dynamics} we proposed a quantisation of the dynamics in a reduced canonical formalism associated with a fixed clock `$T$'. So how can we quantise the same dynamics in other clocks? 

By ``quantisation'' we mean a linear map from phase space functions to linear operators in the Hilbert space that in general takes the following integral form \cite{JPGconf1, JPGconf2}:
\begin{align}\label{IQ}
f(q,p,T)\mapsto \hat{F}_T:=\int_{T=const}\ud q\ud p f(q,p,T)M(q,p),
\end{align}
where $M(q,p)$ is a family of bounded operators in some Hilbert space $\mathcal{H}$ such that
\begin{align}
\int_{T=const}\ud q\ud p~M(q,p)=\id_{\mathcal{H}}~,
\end{align}
and the integration is taken over the constant clock sub-manifolds, i.e. the respective phase space. The integral quantisation (\ref{IQ}) includes all the well-known cases, e.g. the ``canonical prescription", or the Weyl-Wigner quantisation, is obtained for
\begin{equation}
M(q,p)=\mathbf{D}(q,p)2\mathcal{P}\mathbf{D}^{\dagger}(q,p),
\end{equation}
where $\mathcal{P}$ is the parity operator, $\mathbf{D}(q,p)=e^{i(p\hat{Q}-q\hat{P})}$ is the displacement operator and $\hat{Q}$ and $\hat{P}$ are the position and momentum operators on the line \cite{WHcomp}.

For any other choice of clock, say $\bar{T}$, we choose exactly the same family of bounded operators, $M(\cdot,\cdot)$, to define quantisation of the respective phase space,
\begin{align}
f(\bar{q},\bar{p},\bar{T})\mapsto \hat{\bar{F}}_{\bar{T}}:=\int_{\bar{T}=const}\ud \bar{q}\ud \bar{p}f(\bar{q},\bar{p},\bar{T})M(\bar{q},\bar{p}).
\end{align}
In practice, we repeat the quantisation described in the section {\it Quantisation of dynamics} for all other internal clocks in the following way. We promote the basic observables $\bar{q}$ and $\bar{p}$, which are canonical in the clock $\bar{T}$, to the same pair of quantum operators which are given to the basic observables $q$ and $p$, which are canonical in the clock $T$, namely
\begin{align}q \mapsto \hat{Q},~p\mapsto \hat{P}~~\Rightarrow~~\bar{q} \mapsto \hat{Q},~\bar{p}\mapsto \hat{P},\end{align}
where $\hat{Q}, \hat{P}$ are linear operators in $\mathcal{H}$. Upon extending this formal replacement rule to compound observables, we obtain that any $f(q,p)$ and $f(\bar{q},\bar{p})$ are promoted to the same quantum operators, i.e.
\begin{align}f(q,p)\mapsto \hat{F}~~\Rightarrow~~f(\bar{q},\bar{p})\mapsto \hat{F},\end{align}
where $\hat{F}$ is a linear operator on $\mathcal{H}$. This includes the respective Hamiltonians,
\begin{align}\label{qhams}H(q,p)\mapsto \hat{H}~~\Rightarrow~~H(\bar{q},\bar{p})\mapsto \hat{H}.\end{align}
The introduced procedure, based on the formal replacement of the basic observables and clocks in a quantisation map, defines a quantisation of the dynamics in all internal clocks and places it in a unique Hilbert space $\mathcal{H}$. Let us call it ``the extended quantisation procedure". Below we discuss some of its properties.

The observables $f(q,p)$ and $f(\bar{q},\bar{p})$ are promoted to the same linear operator on $\mathcal{H}$, nevertheless they correspond in general to different physical quantities as 
\begin{align}f(q,p)\neq f(\bar{q}(q,p,T),\bar{p}(q,p,T))\end{align}
according to the relation (\ref{particleswitch}) between basic variables and clocks. The only exceptions are constants of motion, which by the virtue of construction given in Eq. (\ref{switch}), have the same dependence on the respective basic variables and  internal clock, i.e. 
\begin{align}I_j(q,p,T)=I_j(\bar{q}(q,p,T),\bar{p}(q,p,T),\bar{T}(q,p,T)).\end{align}
The extended quantisation procedure promotes $I_j$'s to the same quantum operators, $\hat{I}_j$ (or, the same clock-enumerated families of operators $\hat{I}_j(\tau)$, where $\tau\equiv T$ or $\tau\equiv\bar{T}$) for all internal clocks. In other words, the quantum representation of constants of motion is unique. We conclude that any dissimilarities between quantum dynamical operators in different internal clocks are due to different choices of clock rather than other quantisation ambiguities like orderings, choices of basic variables, etc (see also \cite{M}). We provide more details on this idea in Appendix B.

{ Below we describe two methods of comparison of quantum theories based on different clocks and then we apply one of them to the quantum F-L model.}

\subsection{Comparison method A}
Let us focus on dynamical observables. Any dynamical observable $f(q,p)$ never corresponds to the same physical quantity as $f(\bar{q},\bar{p})$ if the relation between clocks $T$ and $\bar{T}$ is non-trivial. Suppose that the observable $f(q,p)$ corresponds to the same physical quantity as the observable 
\begin{align}g(\bar{q},\bar{p})=f(q(\bar{q},\bar{p}),p(\bar{q},\bar{p})),\end{align}
which can be determined from the relation between barred and unbarred variables fixed by Eq. (\ref{switch}) or (\ref{particleswitch}). Hence, a unique physical quantity is promoted to two distinct operators on $\mathcal{H}$: in the first case it is $\hat{F}$ and in the other case it is $\hat{G}$. Since 
\begin{align}\hat{F}\neq\hat{G},\end{align}
they in general have different (generalised) eigenvectors and even different spectra. { We interpret such discrepancies as clear indications of the influence of the choice of clock on the dynamics of a quantum system. To better understand this influence let us assume that $\hat{F}$ and $\hat{G}$ are self-adjoint and that the respective eigenvalue problems read:
\begin{align}\hat{F}|f\rangle=f|f\rangle,~~~\hat{G}|g\rangle=g|g\rangle,\end{align}
where $(|f\rangle,f)$ and $(|g\rangle,g)$ are eigenstates and respective eigenvalues. A state $|\psi\rangle\in\mathcal{H}$, according to the usual interpretation of quantum mechanics, determines the probability of finding a quantum system at a given value of the observable $f(q,p)$,
\begin{align}P_{\psi,T}(f)=|\langle f|\psi\rangle|^2,\end{align}
or, in another clock,
\begin{align}P_{\psi,\bar{T}}(g)=|\langle g|\psi\rangle|^2.\end{align}
Since the operators $\hat{F}$ and $\hat{G}$ are not the same, the probability distribution of a given classical observable in a fixed state $|\psi\rangle$ must depend on the choice of clock, i.e. $$P_{\psi,T}\not\equiv P_{\psi,\bar{T}}.$$ We conclude that the dynamics of a quantum system must be influenced by the choice of clock because the above discrepancy happens only when a quantum state is classically interpreted with a dynamical observable.} 

However, making such an explicit comparison between operators $\hat{F}$ and $\hat{G}$ may be very difficult as it requires determination of the spectrum and eigenvectors of each operator. Fortunately, it is not the only method available. 

\subsection{Comparison method B}
{
In a sense, a complementary method of comparison is to fix a dynamical operator $\hat{F}$, or a set of dynamical operators $\hat{F}_\alpha$, on $\mathcal{H}$ and associate a state $|\psi\rangle\in\mathcal{H}$ with a unique wave function and a unique probability distribution:
\begin{align}\psi(f)=\langle f|\psi\rangle,~~P_{\psi}(f)=|\langle f|\psi\rangle|^2,\end{align}
where $\hat{F}|f\rangle=f|f\rangle$ as previously. However, the operator $\hat{F}$ corresponds to different classical observables in quantum theories based on different clocks. Thus, the wave function $\psi(f)$ and the probability distribution $P_{\psi}(f)$ take different physical interpretations for different clocks as well. In particular, the formula
\begin{align}f=\langle \psi|\hat{F}|\psi\rangle,\end{align}
describes an expectation value, for a fixed state $|\psi\rangle$, of different classical observables for different clocks. Given a set of independent operators $\hat{F}_\alpha$, where $\alpha=1,\dots,2n$ and $2n$ is the dimension of  classical phase space, the set of expectation values
\begin{align}f_\alpha=\langle \psi|\hat{F}_\alpha|\psi\rangle,~~\alpha=1,\dots,2n,\end{align}
assign to a state $|\psi\rangle$ a complete classical state in the phase space. However, for different choices of clock, the classical interpretations of $\hat{F}_\alpha$ and thus, of $f_\alpha$ are different. Therefore, a given state $|\psi\rangle$ is assigned different classical states for different clocks. We interpret those discrepancies as indications of the influence of the choice of clock on the dynamics of a quantum system (although, the state $|\psi\rangle$ is not evolving, the operators $\hat{F}_\alpha$ must be dynamical in order to correspond to different classical observables in different clocks).

An extension of the above comparison method is to let a state $|\psi\rangle$ evolve according to the Schr\"odinger equation based on a unique quantum Hamiltonian (see Eq. (\ref{qhams})). Then, instead of a fixed classical state, we obtain a trajectory in the phase space,
\begin{align}T\mapsto f_\alpha(T)=\langle \psi(T)|\hat{F}_\alpha|\psi(T)\rangle,~~\alpha=1,\dots,2n,\end{align}
which can be interpreted as semiclassical dynamics, that is, quantum dynamics described in terms of classical observables. Naturally, for different choices of clock we expect to get different semiclassical dynamics. 

The above method can be combined with the affine coherent states. The exact quantum dynamics may be replaced with approximate dynamics within a family of coherent states, $T\mapsto |q(T),p(T)\rangle$. We set the basic operators, $\hat{Q}$ and $\hat{P}$, to describe the quantum dynamics by phase space trajectories:
\begin{align}q(T)=\langle q,p|\hat{Q}|q,p\rangle,~~p(T)=\langle q,p|\hat{P}|q,p\rangle\end{align}
The semiclassical trajectories of $q$ and $p$ in the quantum F-L model were derived in the section {\it Phase space portrait} and, in particular, in Eq. (\ref{semiHE}). The comparison involves the reinterpretation of  $\hat{Q}$ and $\hat{P}$ accordingly to the chosen internal clock. Then, the reinterpretation of $q$ and $p$ follows. This simple procedure (i.e. based on the expectation values) is justified by the fact that dissimilarities between semiclassical dynamics in different clocks reflect the most essential dissimilarities between full quantum dynamics.}

\subsection{Comparison of quantum F-L dynamics}

Let us apply the extended comparison method B to the quantum Friedmann-Lemaitre model of the universe. We said in the section {\it Phase space portrait} that the semiclassical dynamics is generated by the semiclassical Hamiltonian which in the phase space $(q,p)$ equipped with the symplectic form $\omega=\ud q\ud p$ reads
\begin{equation}\label{h}H^{sem}(q,p)=p^2+\frac{\hbar^2K}{q^2}.\end{equation}
When we switch to another clock-based formalism, repeat the same quantisation and employ the same coherent states, we obtain (formally) the same semiclassical Hamiltonian
\begin{equation}\label{ht}H^{sem}(\bar{q},\bar{p})=\bar{p}^2+\frac{\hbar^2K}{\bar{q}^2},\end{equation}
which now acts in the phase space $(\bar{q},\bar{p})$ equipped with the symplectic form $\omega=\ud\bar{q}\ud\bar{p}$. The physical meaning of the canonical pairs $({q},{p})$ and $(\bar{q},\bar{p})$ is different. In order to compare the two semiclassical Hamiltonians we need to apply the coordinate relation (\ref{particleswitch}). We find that
\begin{equation}\label{newht}H^{sem}(\bar{q}(q,p),\bar{p}(q,p))=p^2 +\frac{\hbar^2K}{(q+2pD)^2},\end{equation}
which shows that the semiclassical correction responsible for the resolution of singularity depends explicitly on the employed delay function:
\begin{equation}\label{VD}V_D=\frac{\hbar^2K}{(q+2pD)^2}.\end{equation}
We cannot compare the dynamics by solving the equations of motion generated by both the Hamiltonian (\ref{h}) and (\ref{newht}) in the phase space $(q,p)$, because to each of the Hamiltonians there is associated another symplectic form. Nevertheless, the Hamiltonians are conserved along the motion. Therefore, we can compare the contour plots of (\ref{h}) and (\ref{newht}) in $(q,p)$. 

To explain the apparent dependence of the semiclassical correction (\ref{VD}) on the internal clock employed in quantisation (and in derivation of the subsequent semiclassical description) it is sufficient to recall that pseudocanonical transformations preserve the physical trajectories of the {\it classical} motion given by the classical Hamilton equations (\ref{hamFRW}) or (\ref{hamFRWn}). They cannot at the same time preserve the trajectories of the {\it semiclassical} motion and this is why the semiclassical correction is not preserved under these transformations. It is clear that the obtained discrepancies {\it must appear universally} in any quantisation of any gravitational model as long as there are some corrections to the classical motion.

\begin{figure*}[t]
\centering
 \begin{subfigure}{0.3\textwidth}
\includegraphics[width=\textwidth]{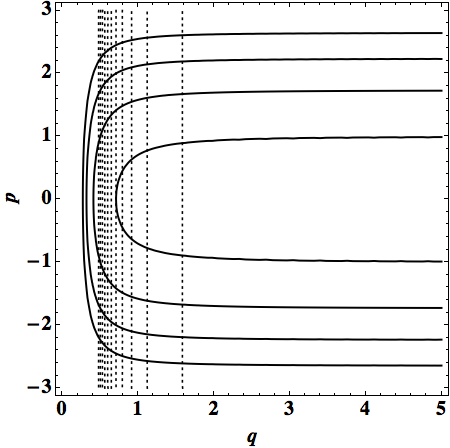}
\caption{}
\label{2}
\end{subfigure}\hspace{2cm}
\begin{subfigure}{0.3\textwidth}
\includegraphics[width=\textwidth]{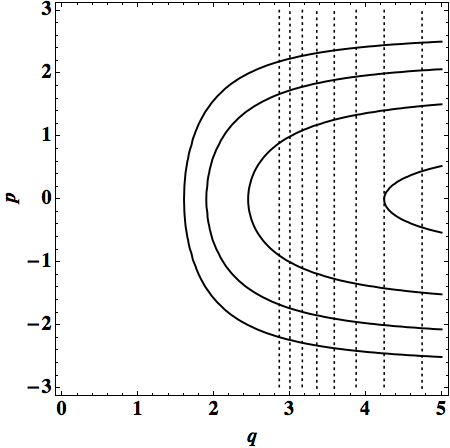}
\caption{}
\label{3}
\end{subfigure}

\vspace{0.6cm}

\begin{subfigure}{0.3\textwidth}
\includegraphics[width=\textwidth]{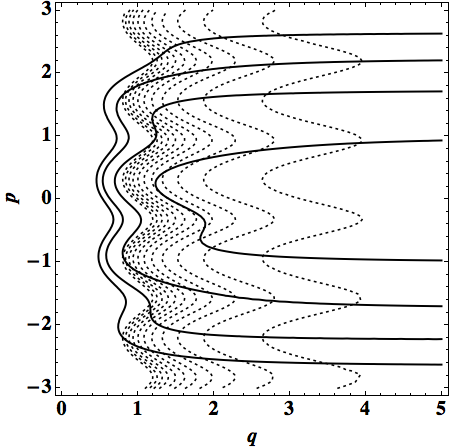}
\caption{}
\label{4}
\end{subfigure}\hspace{2cm}
\begin{subfigure}{0.3\textwidth}
\includegraphics[width=\textwidth]{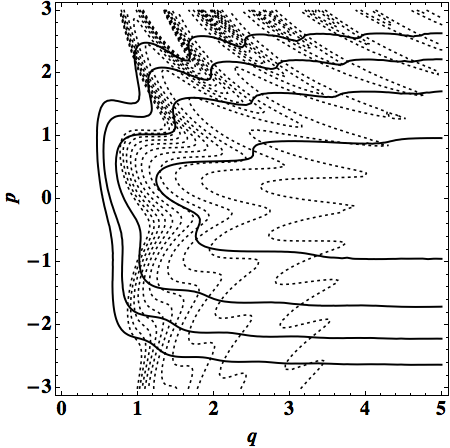}
\caption{}
\label{5}
\end{subfigure}

\caption{The reduced phase space $(q,p)$ is used to compare many semiclassical dynamics. The vertical lines represent repulsive terms $V_D$ which arise from a fixed quantisation with respect to different internal clocks. { The plots demonstrate possible types of dissimilarities between semiclassical dynamics in different clocks. The magnitude of these dissimilarities can be made as large as one wishes by suitable choices of the delay function. We set $K=2$, $q$ and $p$ are given in Planck lengths and Planck masses, respectively.}}

\end{figure*}

We will consider a few examples of clock transformations and effects they have on semiclassical dynamics. In Figs (\ref{2})-(\ref{5}), the semiclassical dynamics are represented by the solid, mostly horizontal lines with a vertical part corresponding to a bounce. The mostly vertical, dotted lines depict the equipotential lines of $V_D$. The values of $H^{sem}$ and $V_D$ used to plot those lines are identical in all the figures.
 \\
\noindent {\it Simple bounce.} We do not redefine the internal clock, we simply put $D=0$. We obtain the dynamics discussed already in \cite{BDGM}. The semiclassical trajectories as well as the equipotential lines are plotted in Fig. (\ref{1}).\\
\noindent {\it Late bounce.} We put $D(q,p)=\frac{1}{2}qp^{-1}$ \cite{footnote}. See Fig. (\ref{2}). The repulsive potential is suppressed and the semiclassical trajectories reach smaller values of $q$ before they bounce. {In fact, the trajectories can be made bounce at volumes as small as one wishes by putting $D(q,p)=\frac{\gamma}{2}qp^{-1}$ with a suitably large value of $\gamma>0$.} We conclude that the moment when quantum effects come to play a dynamical role depends on the choice of internal clock and is not fixed by any scale like Planck scale.\\
\noindent {\it Early bounce.} We put $D(q,p)=-\frac{1}{3}qp^{-1}$. See Fig. (\ref{3}). The repulsive potential is amplified and the semiclassical trajectories bounce at larger values of $q$. {In fact, the trajectories can be made bounce at volumes as large as one wishes by putting $D(q,p)=-\frac{(1-e^{-\gamma})}{2}qp^{-1}$ with a suitably large value of $\gamma>0$.}  This example leads us to the same conclusion as in the previous example.\\
\noindent {\it Multi-bounce.} We put $D(q,p)=q\frac{\sin(5p)}{10p}$. See Fig. (\ref{4}). The semiclassical correction also depends on $p$ and is oscillatory in this variable. Since the bouncing semiclassical trajectories must cross many values of $p$, the oscillatory repulsive term produces many bounces for each trajectory. By adjusting the number of oscillations of the delay function in variable $p$ one may obtain any number of bounces of semiclassical trajectories. This example shows that the number of bounces, or more generally the character of the quantum dynamics, is tied to the choice of internal clock.\\
\noindent {\it Asymmetric bounce.} We put $D(q,p)=q\frac{\sin(3qp)}{10p}e^{p/3}$. See Fig. (\ref{5}). The phase space trajectories are asymmetric in the variable $p$. The universe in the expanding branch undergoes a number of small bumps. Usual semiclassical dynamics found in literature are symmetric about the bounce and are derived with the use of internal clocks that are symmetric about the singularity at the classical level. Such clocks seem in no way better justified than more general ones.

\section{Conclusions}
\label{secVIII}

This work investigates effects of the choice of internal clock on quantisation of dynamics in Hamiltonian constraint systems like general relativity. First, we present some basic elements of the reduced phase space approach to Hamiltonian constraint systems and we introduce the theory of internal clock and the so called pseudocanonical transformations. We apply this approach to quantisation of the Friedmann-Lemaitre model of the universe. We study the quantum dynamics by means of phase space portraits. We show that the classical singularity is avoided in the quantum dynamics due to a quantum repulsive potential. Clock transformations are shown to change the form of the repulsive potential at the quantum level and lead to large modifications in the quantum dynamics of the cosmological model. 

It has been commonly anticipated (see for instance \cite{Th}) that the clock effect should be minor and produce a physically irrelevant ambiguity. This expectation seems to crumble as we show that many physical features of  quantum dynamics can be almost freely altered by switching between clocks. In particular, we show that the concepts like maximal curvature scale, minimal volume or the number of bounces are tied to the choice of internal clock. Concurrently, we note that the most essential feature, i.e. the existence of a bounce which is a smooth transition from contraction to expansion occurs in all internal clocks.

It is very difficult, and perhaps too early, to determine clearly the implications of the obtained result for quantum gravity. Nevertheless, to our mind, the result is too strong to be ignored. It logically follows from the current approach to establishing quantum dynamics for gravitational systems. For the first time we show in concrete examples how the concept of dynamics in quantum gravity based on internal clocks differs from the concept of dynamics in ordinary quantum mechanics based on the absolute time.

Let us point out some issues which could be investigated next. Firstly, we notice that in the present case of the Friedmann-Lemaitre model, a semiclassical trajectory connects fixed contracting and expanding universes irrespectively of the choice of internal clock. It would be interesting to study phase space portraits of higher-dimensional gravitational models to see whether they exhibit this property too. Some steps have been already taken in \cite{M2}. 

Secondly, our result has been obtained within the reduced phase space approach based on the choice of clock made before quantisation. On the other hand, the Dirac approach entails the choice of the internal clock after quantisation. It would be interesting to see whether these two different approaches lead to different results on the issue of internal clocks and to what extent. 

Thirdly, in literature there exist the so called timeless interpretations of quantum gravity (see e.g. \cite{Rovelli}), which assume that the physical predictions cannot depend on the choice of clock. The precise nature of the apparent tension between those interpretations and the reported herein results could be an interesting research subject.

\section*{Acknowledgments} 

This work was supported by Narodowe Centrum Nauki with decision No. DEC-2013/09/D/ST2/03714. I thank Edward Anderson, Jean-Pierre Gazeau and Herv\'e Bergeron for comments on the manuscript. Special thanks to my wife, Elwira.

{\section*{Appendix A: Affine coherent states}

Consider the following transformations of the real line: 
\begin{align}
  \mathbb{R} \ni t \rightarrow (q,p)\cdot t := q^{-1}t + p\in \mathbb{R},
\end{align}
where $(q, p)\in\mathbb{R}_+ \times \mathbb{R}$. The transformations $(q, p)$ form a group with the composition rule:
\begin{align}
(q, p) \circ (q', p') = \left(q q', \dfrac{p'}{q} + p\right)\,.
\end{align}
The neutral element is $(1,0)$ and the inverse of $(q,p)$ is $(q,p)^{-1}= (q^{-1}, -pq^{-1})$.  This group is called the \emph{affine group} of the real line and is denoted by the symbol Aff$_+(\mathbb{R})$. The parameters $q$ and $p$ are called dilation and translation, respectively. The left-invariant measure on the group reads $\ud q \ud p$. Therefore, Aff$_+(\mathbb{R})$ can be identified with the phase space of the F-L model studied in this article and the left-action of Aff$_+(\mathbb{R})$ on itself with a minimal group of canonical transformations of that phase space.

The left-action of Aff$_+(\mathbb{R})$ is unitarily represented in Hilbert spaces parametrised by $\alpha \in \mathbb{R}$:
\begin{align}
\mathcal{H}_\alpha := L^2(\mathbb{R}^*_+, dx/x^{\alpha + 1})\,, 
\end{align}
by operators $U_\alpha(q, p)$ defined as
\begin{align}
U_\alpha(q, p) [\psi(x)] = q^{\alpha/2} e^{ipx} \psi(x/q)\,. 
\end{align}
It has been shown in \cite{gelnai} (see also \cite{aslaklauder}) that $U_\alpha(q, p)$ are irreducible. Throughout the main body of the article we assume $\alpha=-1$.

We apply the action of $U_\alpha(q,p)$ to a fiducial vector $|\psi_0\rangle$ to produce the \emph{affine coherent state}, $| q, p \rangle$,
\begin{align}
\langle x | q, p \rangle = U_\alpha(q, p) [\psi_0(x)] = q^{\alpha/2} e^{ipx} \psi_0(x/q)\, . 
\end{align}
The coherent state $| q, p \rangle$ transforms covariantly under the action of unitary operator $U_\alpha(q', p')$,
\begin{align}U_\alpha(q', p')| q, p \rangle=| q' q, \dfrac{p}{q'} + p' \rangle.\end{align}
At this point, Schur's lemma applies.\\

{\bf Schur's lemma.} \emph{Let $G$ be a group and $U$ its UIR on a vector space $V$. If $M$ is an operator on $V$ such that $U(g) M U(g)^{\dag} = M$ for all $g \in G$, then $M$ is a multiple of the identity on $V$: $M = c\cdot 1$.}\\

It is easy to check that the operator 
\begin{align}M:=\int \ud q \ud p~ | q, p \rangle \langle q, p |,\end{align} satisfies the assumptions of the Schur lemma. Therefore, it follows that
\begin{align}
\int \ud q \ud p~ | q, p \rangle \langle q, p | = c\cdot 1,
\end{align}
where the constant $c$ is finite because the UIR of Aff$_+(\mathbb{R})$ is square integrable.  The affine coherent states are said to resolve the identity w.r.t. the measure $\ud q \ud p/c$.

The affine coherent states are naturally associated with the cosmological phase space, $(q,p)\in \mathbb{R}_+ \times \mathbb{R}$, because they both can be identified with Aff$_+(\mathbb{R})$. This establishes a connection between the classical and quantum level. For example, for any state $|\psi\rangle\in\mathcal{H}_\alpha$, there exists the phase space representation,
\begin{align}\Psi(q,p):=\langle q,p|\psi\rangle,\end{align}
and the phase space probability distribution,
\begin{align}\rho_{\psi}(q,p):=|\Psi(q,p)|^2=|\langle q,p|\psi\rangle|^2,\end{align}
where $\int \rho_{\psi}(q,p) \ud q \ud p/c=1$ for any normalised $|\psi\rangle$ by the virtue of the identity resolution.

If the phase space coordinates are quantised, $(q,p)\mapsto (\hat{Q},\hat{P})$, where $\hat{Q}$ and $\hat{P}$ are operators on $\mathcal{H}_\alpha$, then the classical-quantum connection can be tightened by demanding
\begin{align}\langle q,p|\hat{Q}|q,p\rangle=q,~\langle q,p|\hat{P}|q,p\rangle=p,\end{align}
and each coherent state $|q,p\rangle$ is said to represent the specific classical state $(q,p)$. This opens the door to a semiclassical analysis of quantum dynamics in the spirit of J. Klauder \cite{klauder}. We explain the Klauder approach in Sec. \ref{secVI}.
}

\section*{Appendix B: Uniqueness of quantum representation of Dirac observables}

We notice that in the quantisation defined by Eq. (\ref{IQ}) the internal clock $T$ is external in the sense that the involved measure $\ud q\ud p$  is defined on the phase space rather than the phase space times the time manifold. Analogously, we postulate that for another choice of clock $\bar{T}$, the quantisation in general reads:
\begin{align}\label{IQ2}
f(\bar{q},\bar{p},\bar{T})\mapsto \hat{\bar{F}}_{\bar{T}}:=\int_{\bar{T}=const}\ud \bar{q}\ud \bar{p}f(\bar{q},\bar{p},\bar{T})\bar{M}(\bar{q},\bar{p}),
\end{align}
where $\bar{M}(\bar{q},\bar{p})$ is a family of bounded operators in some Hilbert space $\mathcal{H}$ such that
\begin{align}
\int_{\bar{T}=const}\ud \bar{q}\ud \bar{p}~\bar{M}(\bar{q},\bar{p})=\id_{\mathcal{H}}~.
\end{align}
Notice that the integration takes place along the level sets of the new clock $\bar{T}$. Thus, it is clear that quantisation maps defined in distinct clocks cannot be the same as the integration surfaces $T=const$ and $\bar{T}=const$ are different. Moreover, the application of the coordinate transformation of Eq. (\ref{particleswitch}) to the map (\ref{IQ}) will not transform the canonical measure $\ud q\ud p$ in $T$ into the canonical measure
$\ud \bar{q}\ud \bar{p}$ in $\bar{T}$ as the two measures are inequivalent and the transformation (\ref{particleswitch}) is purely passive. As a result, fixing the quantisation map in one clock and then applying it to all the reduced phase spaces by means of the coordinate transformation (\ref{particleswitch}) would make the initial choice of clock preferred. Therefore, we exclude this procedure and look for another way of relating the maps (\ref{IQ}) and (\ref{IQ2}).

Since a priori $M(q,p)$ and $\bar{M}(\bar{q},\bar{p})$ are unrelated, some dissimilarities between the respective quantum theories can arise due to usual quantisation ambiguities (such as induced orderings) and obscure the effects of the different choice of internal clock. In order to remove the latter we postulate that constants of motions, denoted by $I_n(q,p,T)$ and determined for the F-L model in Eq. (\ref{comFRW}), are given the same quantum representation irrespectively of the choice of internal clock. It makes sense since non-dynamical properties of any system should not depend on the choice of internal clock as its only role is to describe the evolution. According to Eq. (\ref{switch}) {\it any} constant of motion $\mathcal{I}(I_1,\dots, I_{2n})$ is formally a unique function $I(\cdot,\cdot,\cdot)$ of the contact coordinates, i.e. 
\begin{align}\mathcal{I}(I_1,\dots, I_{2n})=I(q,p,T)=I(\bar{q},\bar{p},\bar{T}),\end{align}
irrespectively of the choice of the internal clock. We demand that all classically conserved quantities $\mathcal{I}(I_1,\dots, I_{2n})$ are given the same quantum representation by means of the following equality:
\begin{align}\label{MMeq}
\int_{T=const}\ud q\ud p ~\mathcal{I}(I_1,\dots, I_{2n})~M(q,p)\\ \nonumber
=\int_{\bar{T}=const}\ud \bar{q}\ud \bar{p}~\mathcal{I}(I_1,\dots, I_{2n})~\bar{M}(\bar{q},\bar{p}),
\end{align}
or more explicitly,
\begin{align}
\int_{T=const}\ud q\ud p ~I(q,p,T)~M(q,p)\\ \nonumber
 =\int_{\bar{T}=const}\ud \bar{q}\ud \bar{p}~I(\bar{q},\bar{p},\bar{T})~\bar{M}(\bar{q},\bar{p}),
\end{align}
where we set the values of the respective internal clocks to be equal $T=\bar{T}$. In the $2n$-dimensional phase space equipped with the volume element $\ud q\ud p$, there are $2n$ independent constants of motion $I_j$ and from the arbitrariness of $\mathcal{I}(I_1,\dots, I_{2n})$ in Eq. (\ref{MMeq}) we conclude that
\begin{align}\label{MM}
\forall q=\bar{q},~\forall p=\bar{p},~~ M(q,p)=\bar{M}(\bar{q},\bar{p}),
\end{align}
that is, the family of bounded operators involved in the quantisation must be the same for all phase spaces and all internal clocks in order to satisfy the postulate of a unique quantum representation of constants of motion.


\begin{thebibliography}{99}
\bibitem{footnote}
The delay function $D(q,p)\propto p^{-1}$ is not well-defined at $p=0$. However, this is an admissible choice since there is no classical dynamics at $p=0$ and as a result the following pseudocanonical transformation of canonical variables given by Eq. (\ref{particleswitch}) is well-defined. Regularisation of $p^{-1}$ with a bounded function will not remove any of the discussed effects, it merely introduces extra features around the $p=0$ axis, which depend on the type of applied regularisation and may slightly blur the essential features which are clearly exhibited by the chosen examples.
\end{thebibliography}
\end{document}